\documentclass[12pt]{reportj}
\usepackage{astron} %%Uncomment this if you want to use BibTeX
\usepackage{deluxetablej}     % this line generates a LaTeX warning message
\usepackage{hyperref} %% Links all your references to pages
\makeatletter
\def\@to{to}
\makeatother
\usepackage{times}
\usepackage{graphicx}
\usepackage{tocloft}
\usepackage{fancyheadings}
\emergencystretch=\maxdimen
\usepackage{datetime}
\newdateformat{ddmonthyyyy}{\THEDAY\ \monthname[\THEMONTH]\ \THEYEAR}

\renewcommand\thesection{\arabic{section}}
\renewcommand\thesubsection{\thesection.\arabic{subsection}}

\def\ssection#1{\setcounter{subsection}{0} \refstepcounter{section} \section*{\hbox to \hsize{\large\bf \arabic{section}. #1\hfill }} \addcontentsline{toc}{section}{\arabic{section}. #1}}
\def\ssubsection#1{\setcounter{subsubsection}{0} \refstepcounter{subsection}\subsection*{\hbox to \hsize{\normalsize\bfseries\itshape \arabic{section}.\arabic{subsection} #1\hfill}} \addcontentsline{toc}{subsection}{\arabic{section}.\arabic{subsection} #1}}
\def\ssubsubsection#1{\refstepcounter{subsubsection}\subsection*{\hbox to \hsize{\normalsize\it \arabic{section}.\arabic{subsection}.\arabic{subsubsection} #1\hfill}} \addcontentsline{toc}{subsubsection}{\arabic{section}.\arabic{subsection}.\arabic{subsubsection} #1}}

\def\ssectionstar#1{\section*{\hbox to \hsize{\large\bf #1\hfill}} \addcontentsline{toc}{section}{#1}}
\def\ssubsectionstar#1{\subsection*{\hbox to \hsize{\normalsize\bfseries\itshape #1\hfill}} \addcontentsline{toc}{subsection}{#1}}
\def\ssubsubsectionstar#1{\subsection*{\hbox to \hsize{\normalsize\it  #1\hfill}} \addcontentsline{toc}{subsection}{#1}}

\renewcommand{\cftaftertoctitle}{%
\mbox{}\hfill{\normalfont Page}}
\usepackage{amsmath}

\newcommand*\arcsec{\ensuremath{^{\prime\prime}}}

\newcommand*\farcs{\ensuremath{\overset{\prime\prime}{.}}}

\begin{document}

~\\

% ST Logo in the top left
\vspace{-2.4cm}
\noindent\includegraphics*[width=0.295\linewidth]{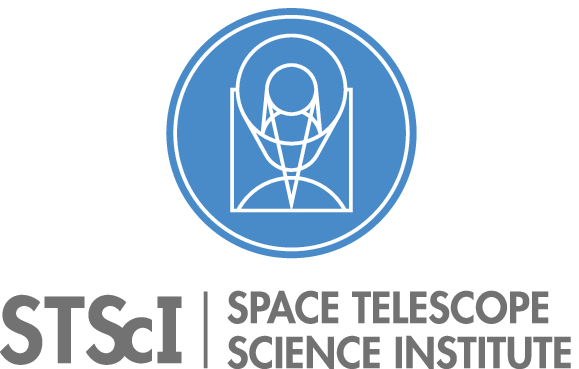}

\vspace{-0.4cm}

\begin{flushright}
    %%% Put the instrument, year, and ISR number here (and also below) %%%
    {\bf Instrument Science Report STIS 2022-05(v1)}
    
    \vspace{1.1cm}
    
    %%% Put ISR Title %%%
    {\bf\Huge Scattered Light in STIS Grating G230LB}
    
    \rule{0.25\linewidth}{0.5pt}
    
    \vspace{0.5cm}
    
    %Put Authors
    Guy Worthey$^1$, Tathagata Pal$^1$, Islam Khan$^1$, Xiang Shi$^1$, Ralph C. Bohlin$^2$
    \linebreak
    \newline
    %Put Author's affiliations
    \footnotesize{$^1$ Washington State University, Pullman, WA 99163\\
                         $^2$ Space Telescope Science Institute, Baltimore, MD \\}
    
    \vspace{0.5cm}
    
    % Date in DD Month YYYY format based on when you compile it
     \ddmonthyyyy\today 
\end{flushright}

% GW added this:
\newcommand{\angstrom}{\mbox{\normalfont\AA}}

\vspace{0.1cm}

%%% ABSTRACT %%%
\noindent\rule{\linewidth}{1.0pt}
\noindent{\bf A{\footnotesize BSTRACT}}

{\it \noindent The G230LB grating used with STIS's CCD detector scatters red light. In red objects, the scattered light mingles with the ultraviolet signal, causing spurious short-wavelength flux and weakening absorption features. Recent calibration observations characterize the scattered light using duplicate observations with the MAMA detector and similar grating G230L. The full two-dimensional spectrum contains little helpful information to mitigate the scattered light problem. For one-dimensional, extracted spectra, the scattered light can be approximately modeled as a ramped pedestal whose amplitude is proportional to the object's $V$-band flux. We present formulae for scattered light corrections. For stars warmer than G0 spectral type, correction is superfluous. Off-slit-center positioning appears not to affect the properties of the scattered light. Therefore, we are able to extrapolate correction formulae for extended objects from the point source formulae. Polynomials for flux corrections due to off-center slit positioning in the 0$\farcs$2 slit are also tabulated. 
}

\vspace{-0.1cm}
\noindent\rule{\linewidth}{1.0pt}

\newpage  % GW put this in to avoid cutting the table of contents in two

%% Table of Contents
%% Need to compile twice to get the page numbers correct
\renewcommand{\cftaftertoctitle}{\thispagestyle{fancy}}
\tableofcontents

%%% MAIN TEXT BELOW %%%

\vspace{-0.3cm}
\ssection{Introduction}\label{sec:Introduction}

Hubble Space Telescope (HST) is influential \cite{2010PASP..122..808A} in part due to its ultraviolet (UV) sensitivity. One of its UV instruments is the Space Telescope Imaging Spectrograph (STIS), which has modes for direct imaging, objective prism spectroscopy, long slit spectroscopy, and echelle spectroscopy \cite{stis_handbook}. There are so many possible permutations of modes, entrance apertures, detectors, filters, and spectral elements that not all of them are supported by Space Telescope Science Institute (STScI) in terms of providing periodic calibration data. One of its popular, supported modes in the UV is low resolution spectroscopy using a CCD detector paired with grating G230LB. This grating/detector pair delivers spectra at approximately 500 km s$^{-1}$ resolution in the range 1600 \AA\ $< \lambda <$ 3100 \AA .

%!!!!!!!!!!!!!!!!!!!!!!!!!!!!!!!!!!!!!!!!!!!!!!!!!!!!!!!!!!!!!!!!!!!!!!!!!!!!!!!!!!!!!!!!!!!!!!!!!!!!!!!!!!!!!!!!!!!!!!!!!!!!!!!!!!!!!!!!!!!!!!!!!!!!!!!!!!!!!!!!!!!!!!!!!!!!!!!!!!!!!!!!!!!!!!!!!!!!!!!!!!!!!!!!!!!!!!!!!!!!!!!!!!!!!!!!!!!!!!!!
%THIS SECTION NEEDS TO GO AFTER THE END OF THE FIRST PAGE AND BEFORE THE END OF THE SECOND PAGE
%Fill in Instrument, Year, and ISR Number and delete "newpage" immediately after this message
%!!!!!!!!!!!!!!!!!!!!!!!!!!!!!!!!!!!!!!!!!!!!!!!!!!!!!!!!!!!!!!!!!!!!!!!!!!!!!!!!!!!!!!!!!!!!!!!!!!!!!!!!!!!!!!!!!!!!!!!!!!!!!!!!!!!!!!!!!!!!!!!!!!!!!!!!!!!!!!!!!!!!!!!!!!!!!!!!!!!!!!!!!!!!!!!!!!!!!!!!!!!!!!!!!!!!!!!!!!!!!!!!!!!!!!!!!!!!!!

\lhead{}
\rhead{}
\cfoot{\rm {\hspace{-1.9cm} Instrument Science Report STIS 2022-05(v1) Page \thepage}}
%%%%%%%%%%%%%%%%

Unfortunately, this grating scatters red light, as noted in $\S$4.1.6 of the STIS Instrument Handbook \cite{stis_handbook}. The stray light mixes with the ultraviolet signal and, for red objects, causes confusion where the two signal sources are comparable. An example of a large project that contains many red point sources is the Next Generation Spectral Library (NGSL) \cite{2004AAS...205.9406G,2008AAS...21116225H,2016ASPC..503..211H}, which targeted several hundred stars of all spectral types in SNAP scheduling mode, in which observations occur in ``leftover'' orbits between larger programs or to shorten long slews. The scattered light should be subtracted from G230LB observations to improve spectrophotometric fidelity. 

To measure the amount of scattered light, we observed three stars previously observed as part of the NGSL program with a different detector/grating pair; the MAMA detector with the G230L grating. The wavelength range and spectral resolution are similar between CCD/G230LB and MAMA/G230L. Unlike the CCD, the MAMA detector cannot ``see'' red photons, and so, while there will presumably be similar amounts of red scattered light in each mode, the MAMA detector is blind to them. We can thus obtain spectra clean of scattered light and compare them to the contaminated versions in order to isolate the scattered component. If the scattered light is amenable to modeling, the correction can be applied to all programs in which red objects were observed with G230LB.

The light path through the STIS instrument comes after two reflections from the primary and secondary mirrors of the telescope. The spectrograph inserts three more reflections before the long slit entrance pupil and the grating disperser \cite{stis_handbook}. Two more reflections occur after dispersion, and then the light falls on the detector, for a grand total of seven reflections, a long slit aperture, and a grating. The instrument is 285$\arcsec$ off-center from the focal axis, which causes the point spread function (PSF) to become asymmetric. Red wavelength PSFs show Airy disks, but for UV observations the first Airy null contracts to lie inside a lumpy PSF core. The PSF can be time-variable due to thermal expansion and contraction between optical elements, focus ``breathing'' effects (ISRs 2018-06, 2017-01), and also variations in charge diffusion within the detectors. 

We describe the observations in more detail in $\S$2. We show the outcomes and provide various correction formulae in $\S$3 and conclude in $\S$4.

%% SECTION
\vspace{-0.3cm}
\ssection{Observations}\label{sec:obs}

During program GO 16188, we targeted K0 V star BD~+41~3306, K4 III star HD 13520, and M1 III star HD 102212 in order to test over a color range. Each star consumed two orbits, one dedicated to CCD/G230LB observations and one dedicated to MAMA/G230L observations. The observation dates and exposure times are given in Table \ref{obslog}. Because the stars are bright, exposure times could be short. We used the leftover time within an orbit to explore the effects of placing the star off the centerline of the slit. During HST's normal observing cadence, a point source is ``acquired'' with nominal error of $0\farcs01$ and then, for observations with narrow slits, ``peaked up'' with nominal error 5\% of whatever slit width is being used (typically, the 0$\farcs$1 slit; 5\% of 0$\farcs$1 is $0\farcs005$). The NGSL program, with its 0$\farcs$2-wide slit, skipped the peak-up because of time constraints during orbit-packing. Stars can therefore be displaced relative to slit centerline by a nominal $0\farcs01$. If the slit is changed, introduced positional change is only $0\farcs0025$. If the grating is swapped, a positional shift of $\leq \pm$3 MAMA pixels, or at a plate scale of $0\farcs025$ per pixel, a shift of $\leq \pm 0\farcs075$ (STIS Instrument Handbook $\S$3.3), but this only affects where the light lands on the detector, not how much light gets through the slit.

In order to be of maximum utility for the NGSL project, we also adopted a slit width of 0$\farcs$2, and then varied the star placement in 0$\farcs$05 increments from center. Because of the asymmetric PSF of the off-axis STIS instrument, the malcentering causes wavelength-dependent spectrophotometric errors, and the errors from a shift to shorter wavelengths do not perfectly match those from a shift to longer wavelengths.

   \begin{table}
      \caption[]{Targets and exposure times in seconds.}
         \label{obslog}
     $$ 
         \begin{array}{lcccccc}
            \hline
            \noalign{\smallskip}
            {\rm Target}     & {\rm Sp.} & V & V-K & {\rm Date} &  {\rm CCD}  & {\rm MAMA}\\
               &  & (mag) & (mag) &  &  (s)  & (s) \\
            \noalign{\smallskip}
            \hline
            \noalign{\smallskip}
            {\rm BD}\ +41\ 3306 & {\rm K0\ V}   & 8.86 & 2.16 & 2021\ {\rm Feb}\ 27  & 240 & 300  \\
            {\rm HD}\ 13520     & {\rm K4\ III} & 4.83 & 3.49 & 2021\ {\rm Feb}\ 07  & 160 & 300  \\
            {\rm HD}\ 102212    & {\rm M1\ III} & 4.04 & 3.97 & 2021\ {\rm Mar}\ 02  & 120 & 180  \\
            \noalign{\smallskip}
            \hline
         \end{array}
     $$ 
   \end{table}

The observations of Feb/Mar 2021 included the primary pointing peak-ups and wavelength calibration exposures. All observation attempts succeeded, and the data was reduced via the standard pipeline currently employed by the Mikulski Archive for Space Telescopes \cite{data_handbook}. 

\begin{figure}
   \centering
   \includegraphics[width=0.8\columnwidth]{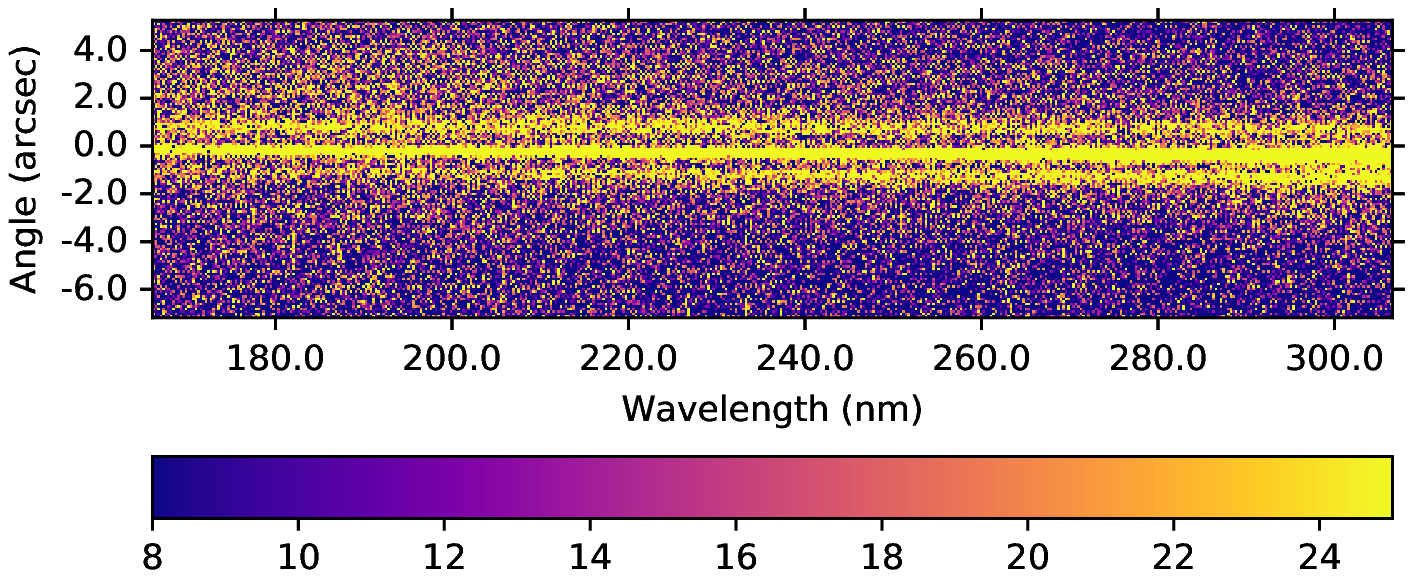}

   \includegraphics[width=0.8\columnwidth]{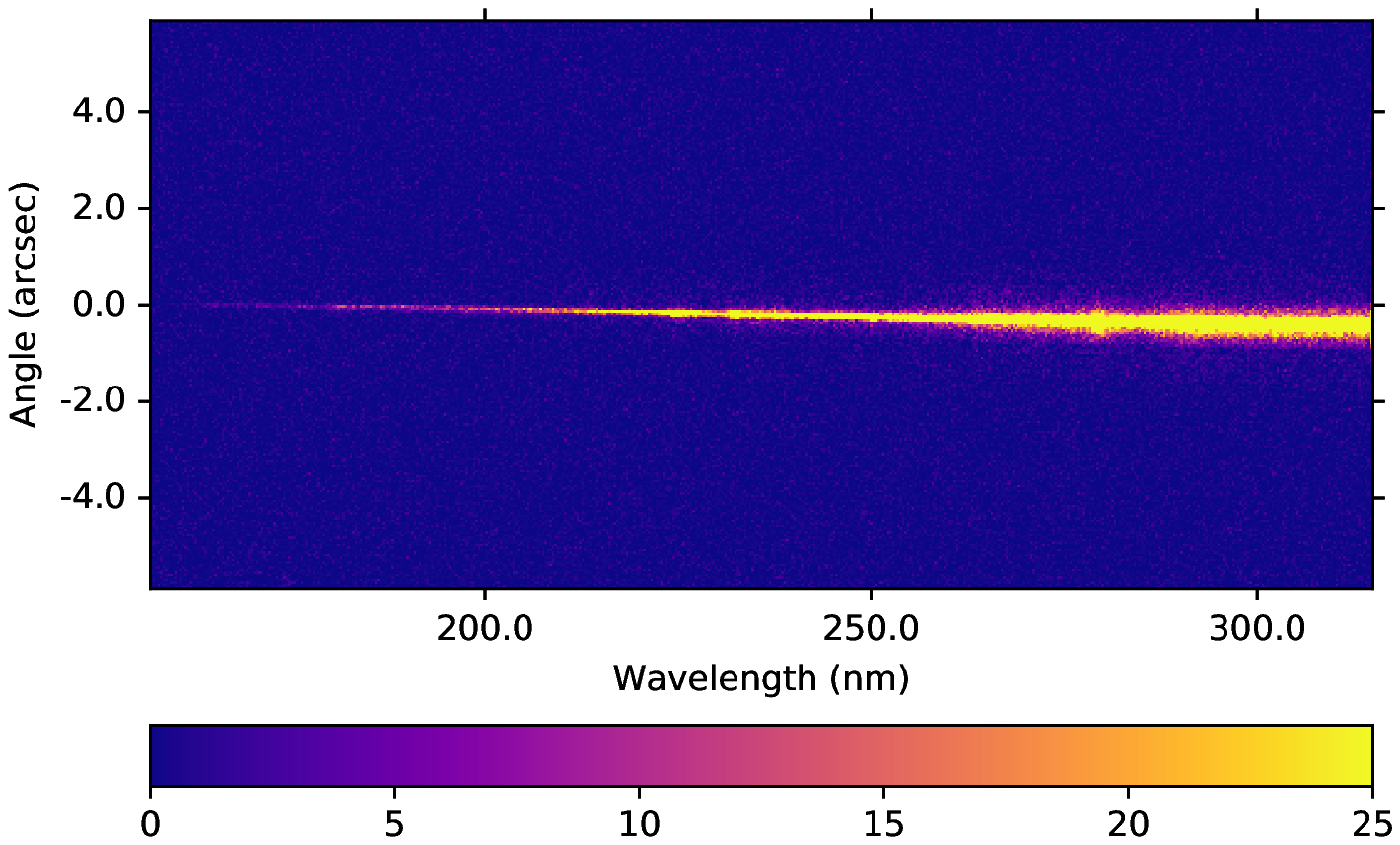}

   \caption{Two-dimensional spectra for HD 13520. The CCD/G230LB image (top) and the MAMA/G230L image (bottom) are stretched to show faint substructure. In the CCD image, the left (blue) half of the spectrum is essentially all scattered light and ``railroad track'' spectra can be seen. In the MAMA version, no scattered light is obvious. Color maps are in units of counts s$^{-1}$.
              }
         \label{ccd2d}
   \end{figure}

The hope of isolating the scattered light also succeeded. Figure \ref{ccd2d} (top) shows a CCD frame for HD 13520. The spatial dimension is vertical and the spectral dimension is horizontal with red to the right. Faint parallel spectra are seen both above and below the main trace, a fact that should be noted when choosing windows over which to define sky for spectral extraction. Section 13.7.4 of the STIS Instrument Handbook \cite{stis_handbook} and also ISR 98-24 mention these ``railroad tracks.'' Almost all the light in the left half of the CCD (top) trace of Fig. \ref{ccd2d} is scattered red light.
%----------------------------------------------------------------- 
%-----------------------------------------------------------------

Fig. \ref{ccd2d} (bottom) shows the same spectrum as seen with the MAMA detector through the G230L grating. The spectral scale is slightly compressed in comparison, and the spectral trace tilts more in the spatial dimension, but there is no evidence of scattered light. This shows that the scattered light is primarily of longer wavelength.

\begin{figure}
   \centering
   \includegraphics[width=\columnwidth]{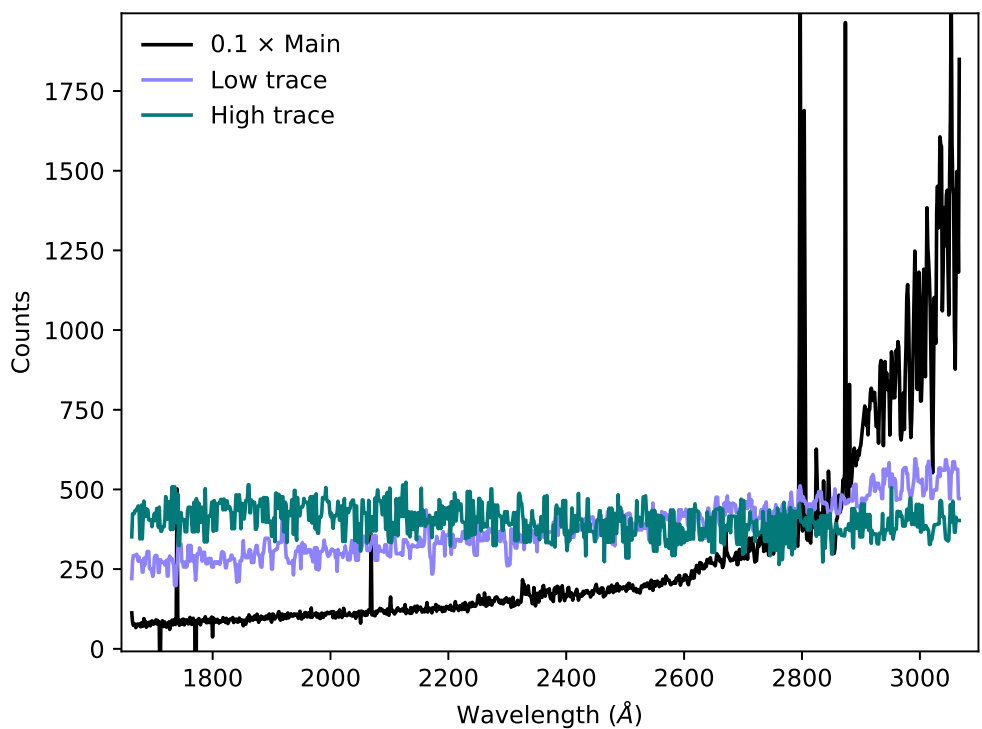}
      \caption{Extracted railroad track spectra (purple for the spectrum at lesser $y$, bluegreen for the spectrum at greater $y$), along with the main stellar spectrum (black). This is HD~13520, corresponding to the CCD image of Fig.~\ref{ccd2d}. 
              }
         \label{ghost2}
   \end{figure}

The railroad track spectra in the CCD image contain substantial amounts of flux. We extracted the flanking parallel spectra themselves and compared them with the main trace in Fig. \ref{ghost2}. A contribution from starlight may exist in the lower trace (that is, at lesser $y$ pixel), seen at longer wavelengths as enhanced flux. This perhaps implies some specular reflection in a direction perpendicular to the dispersion direction. However, some prominent stellar spectral features do not appear in the lower spectrum, and also the lower track also resembles the overall tilted-pedestal profile of the scattered light discussed below. The upper railroad track spectrum appears nearly constant as a function of wavelength, implying a pure scattering origin. Furthermore, the BD+41 3306 observations show no railroad track spectra. Since BD+41 3306 is warmer, it has comparatively less scattered light, and the fact that the railroad track spectra disappear implies that they are additional scattered light artifacts.

%----------------------------------------------------------------- 
%-----------------------------------------------------------------

The railroad track spectra have a lower count rate than the main trace of scattered light, but  different spectral tilts compared to each other and the main trace. For point sources, sky extraction zones can be chosen to exclude the extra spectra, so they do not add to the scattered light problem. If they are extracted, the extra information contained in them does not help disentangle scattered light and signal in the main trace. There appears no particular advantage to considering the extra information contained in the full 2d image when modeling the scattered light within the main trace.

\begin{figure}
   \centering
   \includegraphics[width=\columnwidth]{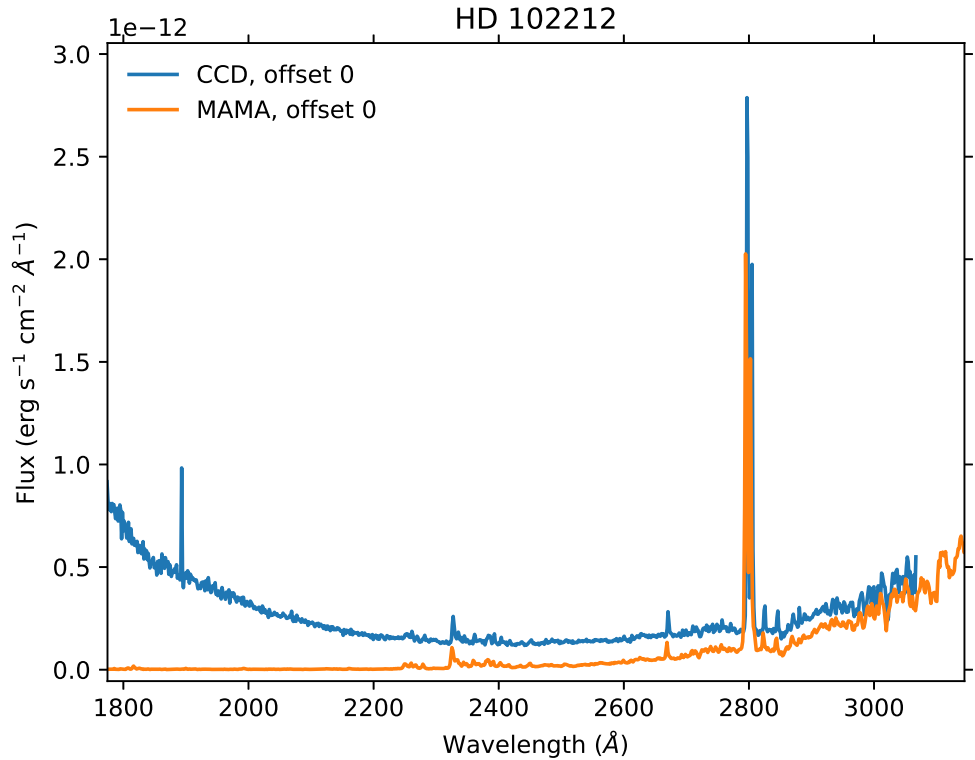}
      \caption{Example spectra after fluxing, from CCD/G230LB (blue) or MAMA/G230L (orange). Flux-correction amplifies small amounts of scattered light at the short wavelength end of the spectrum to resemble the continuum of a hot source.
              }
         \label{somespectra}
   \end{figure}

Extracted to one-dimensional spectra and then fluxed, the amplified scattered light mimics the profile one might expect if (all) cool stars all have a hot companion, as shown in Fig. \ref{somespectra}. Of course, the signal is not astrophysical, but instrumental. We now go on to model the scattered light and also attempt to characterize the effects of improperly placing the star relative to the centerline of the $0\farcs2$-wide long slit.
%----------------------------------------------------------------- 
%-----------------------------------------------------------------

%% SECTION
\vspace{-0.3cm}
\ssection{Results}\label{sec:results}

%% Subsection
\vspace{-0.3cm}
\ssubsection{Scattered light}\label{sec:scatlight}

\begin{figure}
   \centering
   \includegraphics[width=\columnwidth]{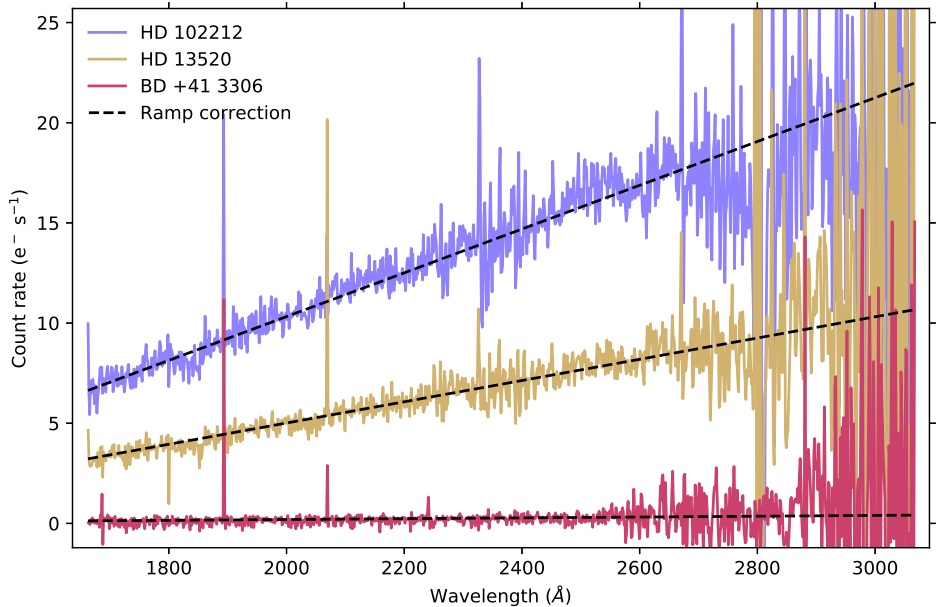}
      \caption{Inferred scattered light distribution for our three program stars (solid colored lines) along with ramp models (dashed lines) for correction.
              }
         \label{ramps}
   \end{figure}

Our three stars have different surface temperatures and spectral types (M1, K4, and K0 for HD 102212, HD 13520, and BD~+41~3660, respectively). Spectral extractions used a 11-pixel window recommended for best fluxing (for bright targets; Lindler, private communication) and subtracted background $\pm$30 pixels away from the main trace in 5-pixel strips in order to miss the railroad track spectra. These parameters differ slightly from the \textit{stistools x1d} default). All count rates we quote are \textit{after} CTE correction (default version in the present case). Subtracting the MAMA-observed 1-d spectrum from the CCD-observed spectrum yields a scattered light profile, shown in Fig. \ref{ramps}. No obvious spectral features appear in the scattered light. Both cooler stars have prominent chromospheric emission from the Mg II doublet ($\lambda \approx$ 2800), but aside from that spectral window residuals from a line fit are uncorrelated, implying that the stellar spectrum provides no power to the scattered light signal. The basic character of the scattered light resembles a ramp, with scattered light concentrated at the long-wavelength end of the spectrum.

%----------------------------------------------------------------- 
%-----------------------------------------------------------------

The simplicity of the scattered light spectrum (a linear ramp) argues against hypotheses that involve specular reflections. On the other hand, isotropic diffuse reflections are also ruled out because the light would fall over large portions of the detector. The scattered light is spatially confined to the beam except for the railroad track images seen in Fig. \ref{ccd2d} but widely dispersed along the spectral dimension. This reasonably indicates that it is the dispersing element that causes the scattering, in agreement with the opinion of Dashevsky \& Caldwell \cite*{2000AAS...196.3211D}. The two gratings we exploit, G230L and G230LB, one might imagine were manufactured identically, but they are not \cite{1998PASP..110.1183W}. G230L has 103.7 grooves mm$^{-1}$ and a blaze angle of 0.74$^{\circ}$ while G230LB has 246.2 grooves mm$^{-1}$ and a blaze angle of 1.41$^{\circ}$ in order to compensate for the fact that the MAMA detector is half the physical size of the CCD. The result is that the MAMA and CCD see approximately the same swath of spectrum.

We model the scattered light with a sloping pedestal that is a step more complicated that the flat pedestal used by Dashevsky \& Caldwell \cite*{2000AAS...196.3211D} but very similar to the ramp used by Lindler \& Heap \cite*{MASTNGSL}.  If the scattered light at 2000\AA\ is given by $K_0$, the correction function

$$ S(\lambda) = K_0 [1.0 + 0.00104*(\lambda - 2000\ \angstrom )]$$

\noindent fits our stars well as seen in Fig. \ref{ramps}. $S(\lambda)$ and $K_0$ are in units of electrons per second, like a pipeline-reduced spectrum before fluxing. To get units of electrons per second, multiply the NET counts returned by \textit{stistools x1d} by the ATODGAIN from the header.

\begin{figure}
   \centering
   \includegraphics[width=\columnwidth]{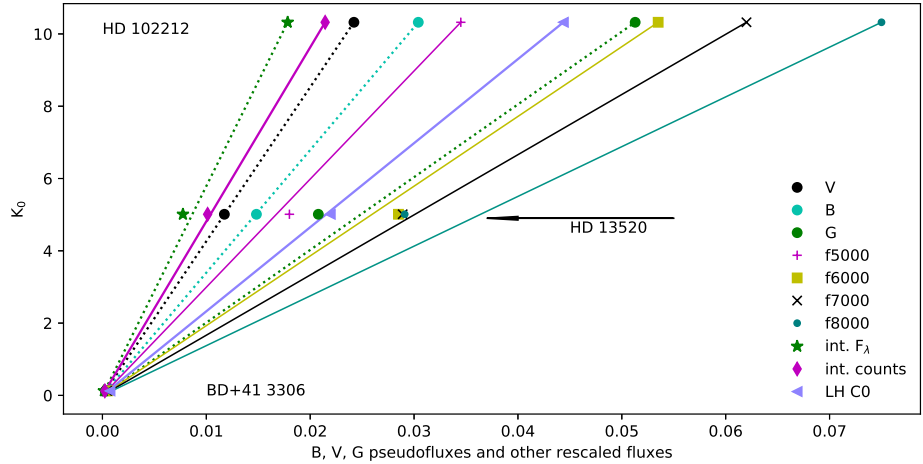}
      \caption{Amplitude of the scattered light $K_0$ versus various potentially predictive fluxes. The x-axis plots monochromatic fluxes (in erg s$^{-1}$ cm$^{-2}$ \AA$^{-1}$ scaled up by $5 \times 10^8$), pseudofluxes (for $q =$ B, V, or G, where a pseudoflux is $10^{-0.4 q}$, and the B pseudoflux is multiplied by an additional factor of 5 for display purposes), estimated total count rate $C$ (scaled by $10^{-9}$), integrated $F_\lambda$ over the CCD bandwidth (scaled by $2.5\times 10^4$), and $C0$ (scaled by 0.075). The $K_0$ amplitudes of scattered light at 2000\AA\ are 10.32 counts s$^{-1}$ for HD 102212, 5.01 counts s$^{-1}$ for HD 13520, and 0.194 counts s$^{-1}$ for BD~+41~3306. Lines are not fits, but assume that HD 102212 is the fiducial. One can then examine the predictions for HD 13520 (indicated by the black arrow) to see if the linear model matches. Flux at $B$, flux at $V$, and integrated counts $C$ land near the line, indicating consistency. 
              }
         \label{chromatic}
   \end{figure}

This correction must be applied before transforming from count rate to flux. The amount of correction ($K_0$) could be predicted in a variety of ways. Brighter stars will generate more scattered light, we expect proportionally. But at which wavelength should the brightness be assessed? For our observations, we chose stars of three different colors in order to address this question. An indication of the results is shown in Figure \ref{chromatic}, where we draft various fluxes as indicators of $K_0$ and see if we can find consistency for a formula of the type 

$$ K_0 = A \times F $$

\noindent where $F$ is a flux and $A$ is a proportionality constant. If we choose the correct sort of $F$, then the model will fit stars of different surface temperatures. Figure \ref{chromatic} is a testing ground for various candidate fluxes. We try broad-band photometric filters B, V, and \textit{Gaia} G, transformed to pseudofluxes via $P_X = 10^{-0.4 X}$, where $X$ is a standard magnitude and $P_X$ is a pseudoflux. We also test monochromatic fluxes. Because our stars exist in the Lindler \& Heap \cite*{MASTNGSL} collection of spectra, we have monochromatic fluxes throughout the UV and optical to choose from. Finally, we take the flux integrated over the NGSL wavelength span and also a fictitious integrated total count rate. This comes in two varieties, both based on a CCD response curve from pre-launch data (Lindler, private communication) sampled at 10-\AA\ invervals that is scaled to match the count-rate to flux conversion in the G230LB wavelength interval. In the first, the stellar spectrum is multiplied by this curve between 2000\AA\ and 10000\AA\ at the tabulated 10-\AA\ intervals, and summed to get a value $C$ proportional to the total count rate as if the CCD could catch the whole spectrum coming through G230LB. The second variety is $C0$, computed the same way except the CCD response curve is first multiplied by $\lambda^{-3}$ with $\lambda$ in \AA\ units, as suggested by Lindler \& Heap \cite*{MASTNGSL}. Lindler \& Heap's correction appears to have worked well judging by their final NGSL spectra, but we were not able to reproduce its zeropoint from the information published, even with the CCD response curve from Lindler. Here, we derive our own (Table \ref{ktab}).

%----------------------------------------------------------------- 
%-----------------------------------------------------------------

   Figure \ref{chromatic} shows how HD 13520 fares, but of course all three stars should be considered. That said, the $K_0$ value for BD~+41~3306 is uncertain due to photon noise, and it does not contribute meaningfully. The mean of the two remaining estimates for the slope $A$ yielded the results in Table \ref{ktab}, along with the absolute value of their difference, expressed as a fraction, to give a crude estimate of quality. This table gives formulae for predicting $K_0$ given various fluxes, in order of preference. We find that the $V$ or $B$ magnitudes of the star are the most consistent predictors for $K_0$, the scattered light count rate at $\lambda$2000, for our three-star sample. This is convenient due to the ubiquity of a $V$ magnitude. The integrated count rate parameters and some monochromatic fluxes are also acceptable predictors, but these can only be used when both UV and optical spectra have been obtained. From these considerations, the ``effective wavelength'' of the scattered light can be considered to be near 5000\AA .

If a neutral density filter was used for the G230LB observations, then $K_0$ should be attenuated by the same factor. 

\begin{table}
\caption{Formulae for $K_0$, with fractional absolute deviations for slope $A$.}             % title of Table
\label{ktab}      % is used to refer this table in the text
\centering                          % used for centering table
\begin{tabular}{lll}        % centered columns (4 columns)
\hline\hline                 % inserts double horizontal lines
 title & formula & $\Delta A / A$ \\    % table heading 
\hline                        % inserts single horizontal line
   $K_0 =$ & $426 \times 10^{-0.4\ V} $ & 0.00 \\      % body of the table
   $K_0 =$ & $1697 \times 10^{-0.4\ B} $ &  0.00\\ 
   $K_0 =$ & $9.64 \times 10^4 \times\ C0$ & 0.02  \\
   $K_0 =$ & $4.81 \times 10^{-7}  \times\ C$ & 0.03  \\
   $K_0 =$ & $166 \times 5 \times 10^8 \times F_{7000}$ & 0.04 \\
   $K_0 =$ & $299 \times 5 \times 10^8 \times F_{5000}$ & 0.07 \\
   $K_0 =$ & $193 \times 5 \times 10^8 \times F_{6000}$ & 0.09 \\
   $K_0 =$ & $1474 \times 5 \times 10^8 \times F_{4000}$ & 0.12 \\
   $K_0 =$ & $1.44 \times 10^7 \times\ F_\lambda$ & 0.12 \\
   $K_0 =$ & $201 \times 10^{-0.4\ G} $ & 0.20 \\ 
   $K_0 =$ & $138 \times 5 \times 10^8 \times F_{8000}$ & 0.25 \\
\hline    
    
\end{tabular}
\end{table}

Moving to an even more precise predictor of $K_0$, for example by using a magnitude \textit{and} a color, should not be attempted until more stars at more temperatures are observed. 

\begin{figure}
\includegraphics[width=0.75\columnwidth]{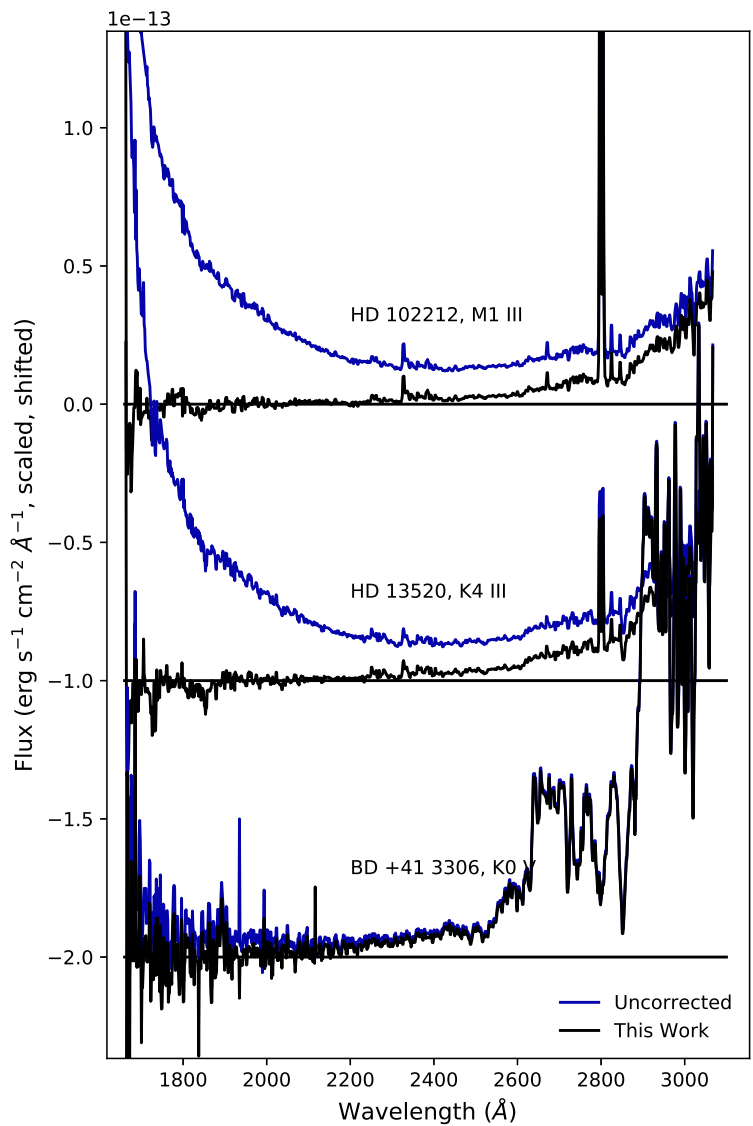}
\caption{Uncorrected (blue) and scattered-light-corrected (black) spectra for our program stars. Because these are our program stars, the correction is constrained to fit optimally within the noise.  \label{fig_flux5}}
\end{figure}

\begin{figure}
\includegraphics[width=0.5\columnwidth]{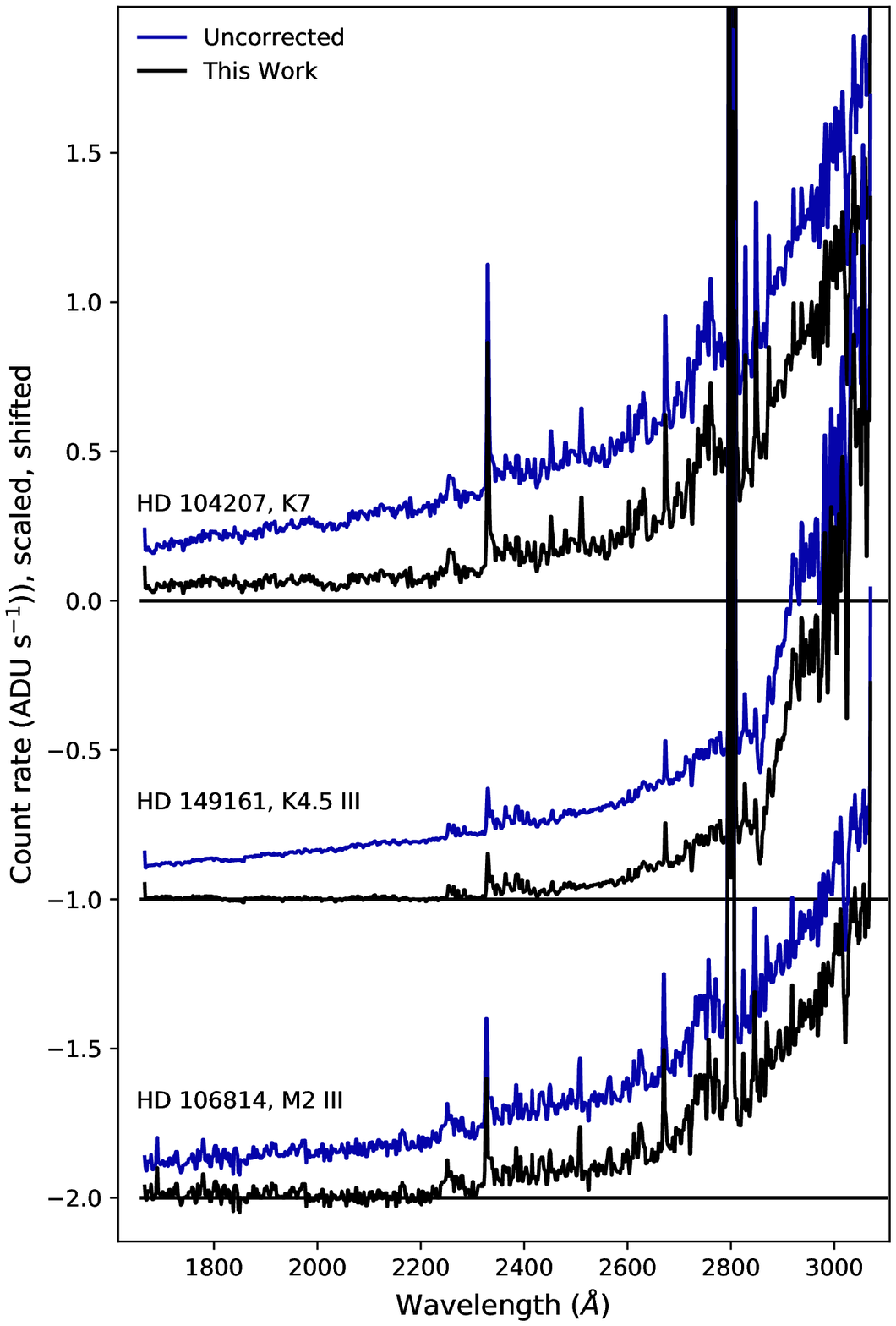}\includegraphics[width=0.5\columnwidth]{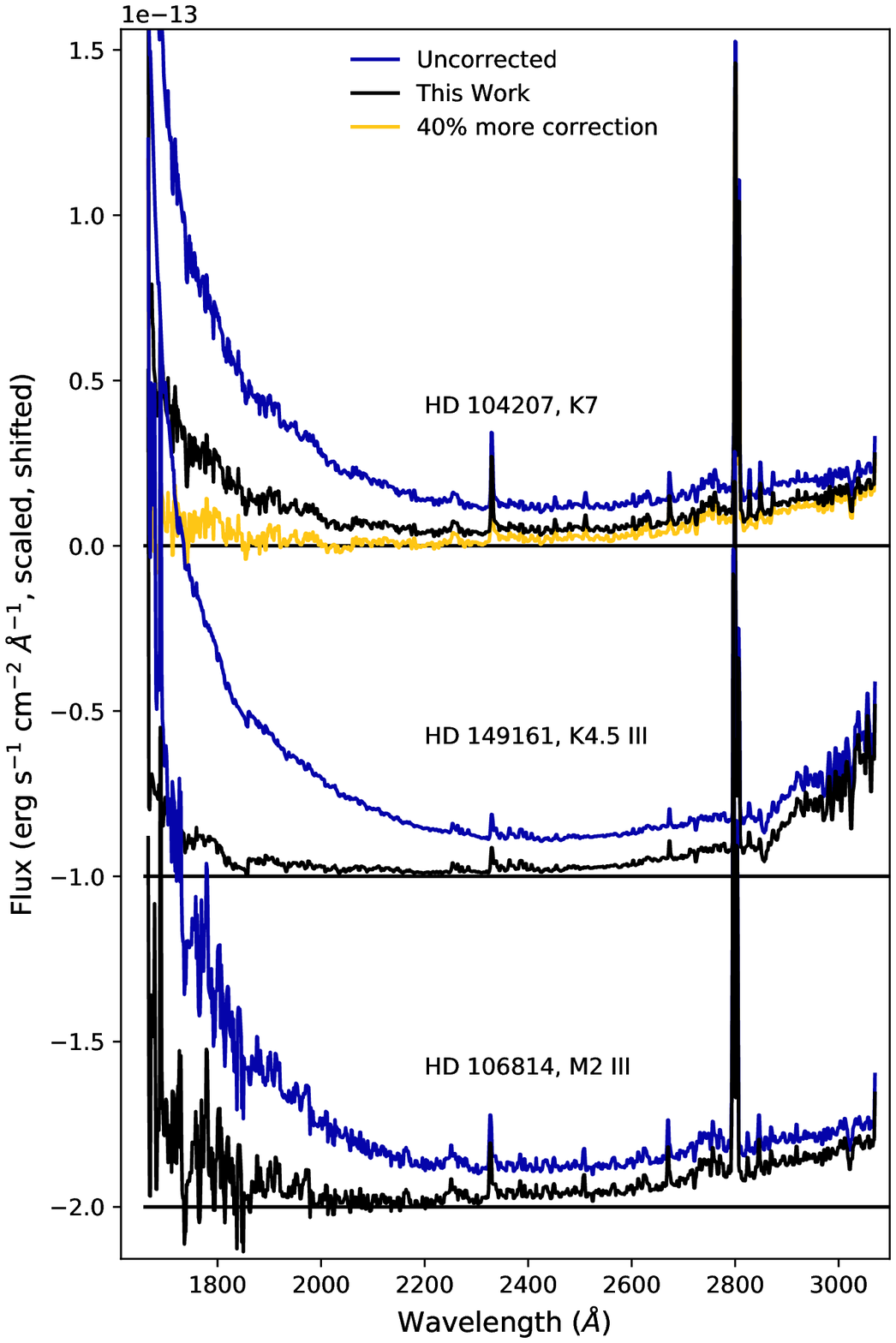}
\caption{Uncorrected (blue) and scattered-light-corrected (black) spectra for stars from program GO 9088. The left panel shows spectra in units of count rate, while the right panel shows the same spectra after conversion to flux units. HD 104207 evidently requires more correction (yellow) than we have derived. \label{fig_p1}}
\end{figure}

\begin{figure}
\includegraphics[width=0.5\columnwidth]{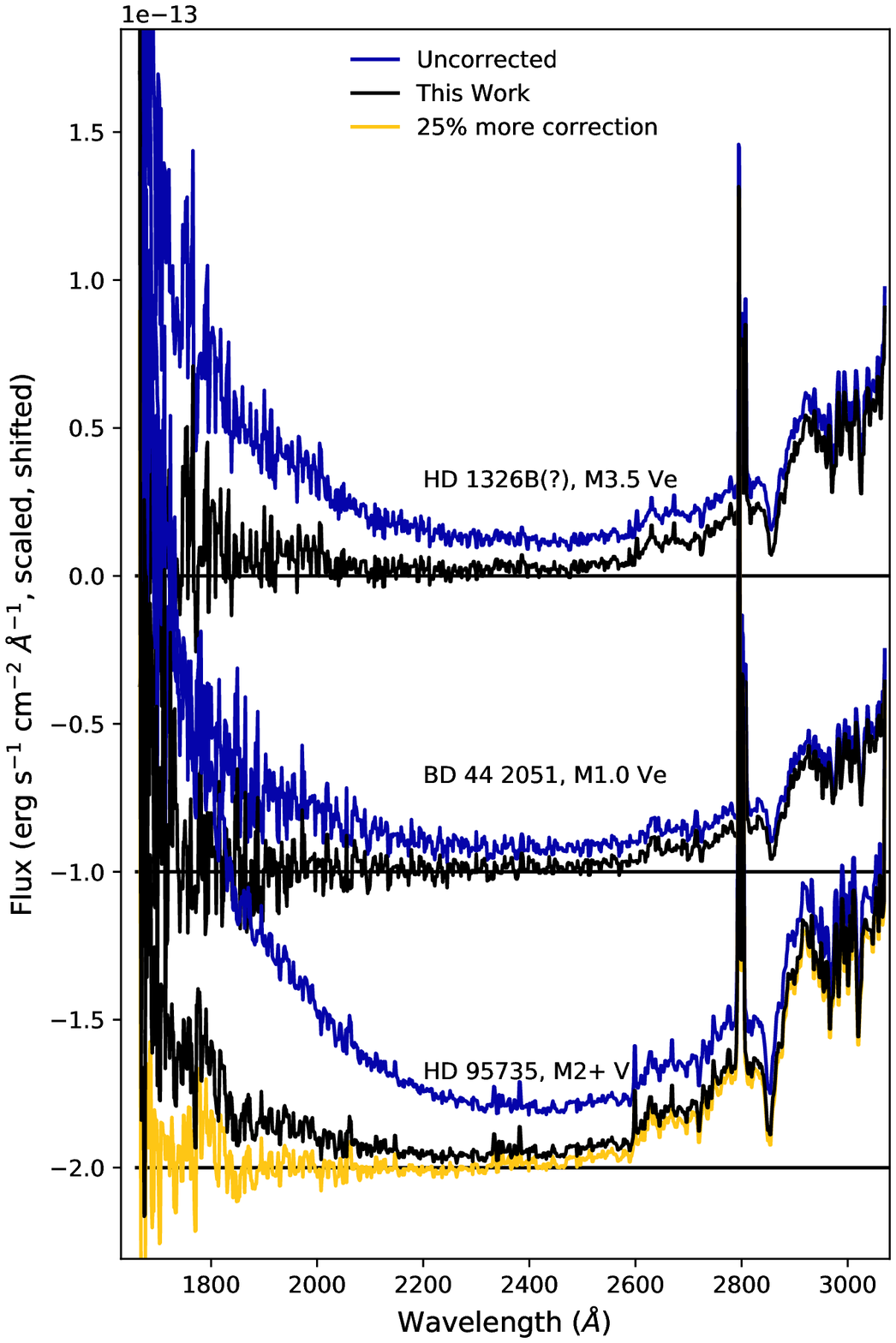}\includegraphics[width=0.5\columnwidth]{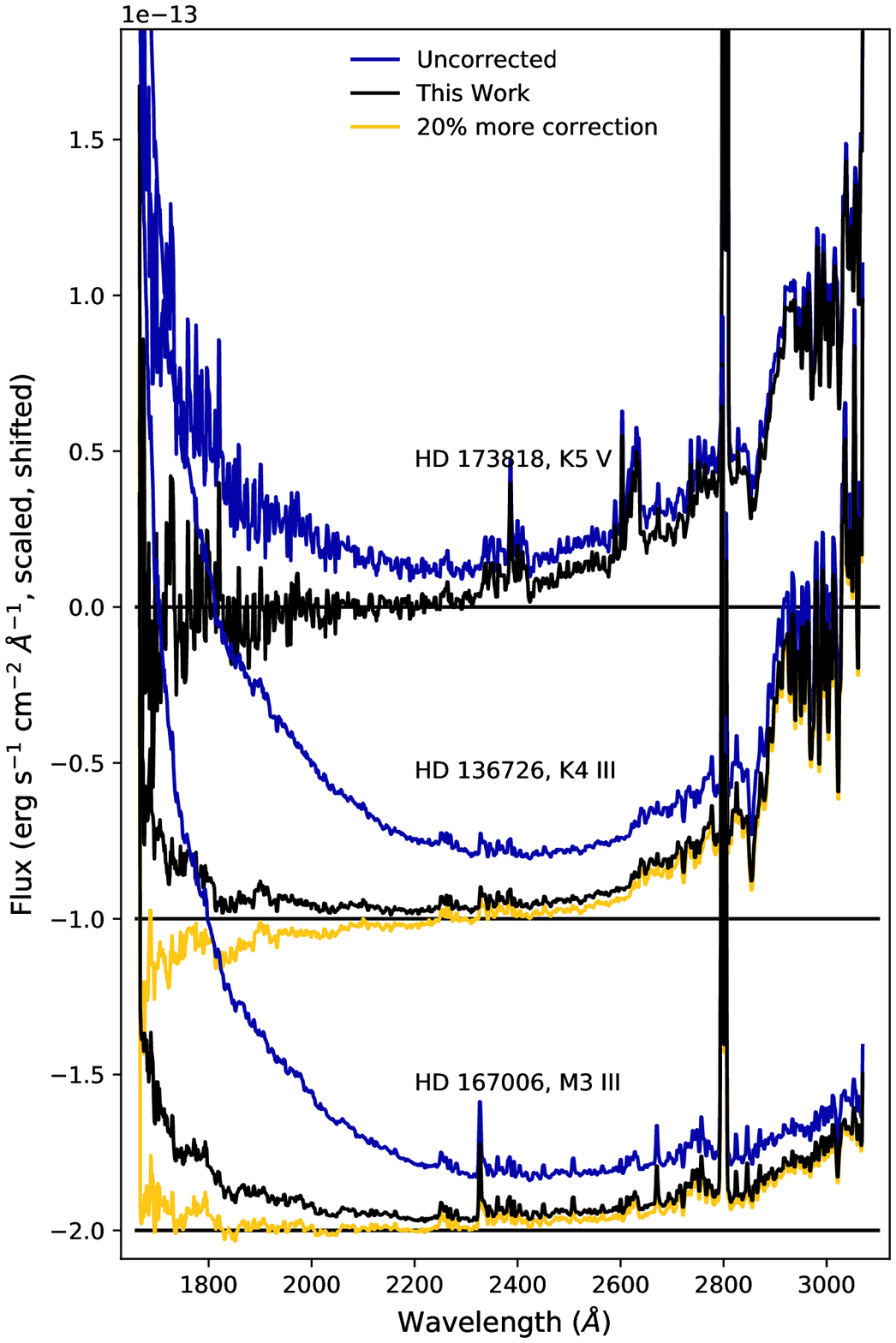}
\caption{Uncorrected (blue) and scattered-light-corrected (black) spectra for stars from program GO 9088. Due to consistency between count rates and expected brightness, we are of the opinion that the observation that targeted HD 1326B actually observed its brighter M2 V neighbor HD 1326, and we show the correction for the latter case. Some example modified corrections (yellow) illustrate the effects of undercorrection and overcorrection. \label{fig_p2}}
\end{figure}

\begin{figure}
\includegraphics[width=0.5\columnwidth]{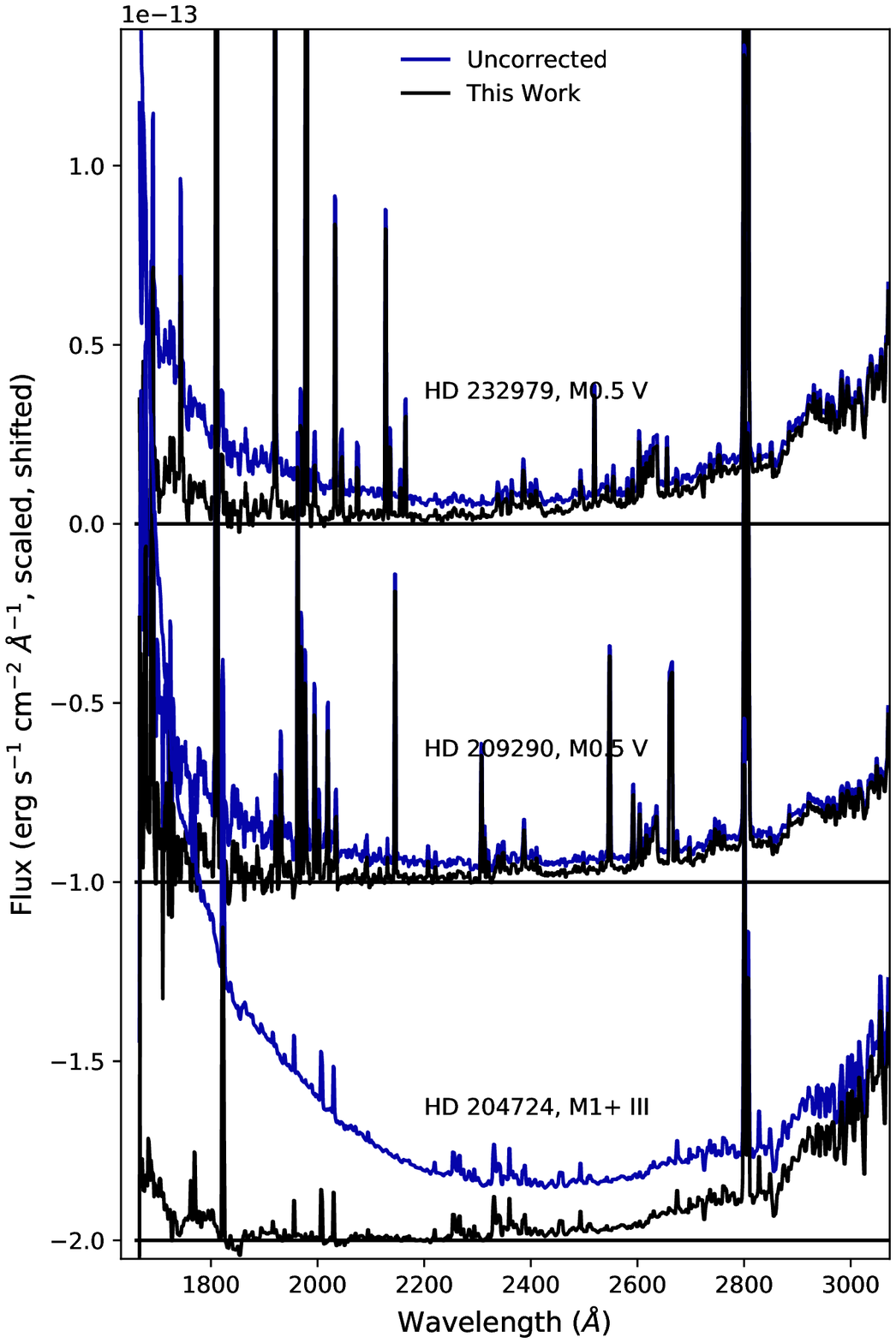}\includegraphics[width=0.5\columnwidth]{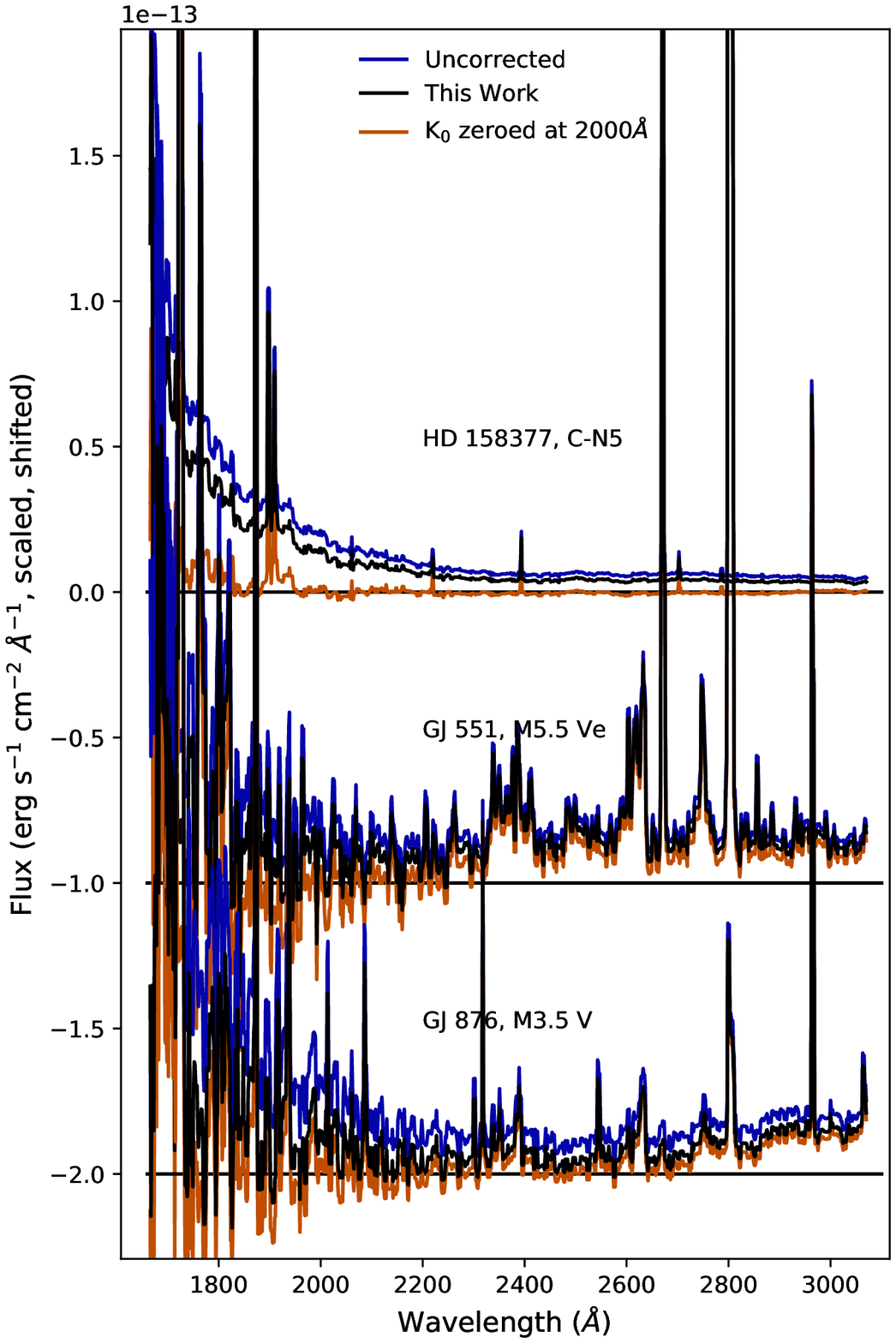}
\caption{Uncorrected (blue) and scattered-light-corrected (black) spectra for cool stars from program GO 13776 after fluxing. The right panels also show the effects of measuring $K_0$ directly from the count-rate spectrum at 2000\AA\ (brown), justified when the stars are cool enough to have negligible photospheric flux at 2000\AA. HD 158377 is a telling case because it is a carbon star with near-zero UV flux, but also an extremely red optical spectrum much different from M stars. It is instructive that this star is undercorrected, but the five non-peculiar stars are well-corrected by the untweaked formula. GJ 551 is Proxima Centauri, a flare star. \label{fig_p3}}
\end{figure}

Circular confirmation that the scattered light correction we derive applies well to the stars from which we derived it appears in Figure \ref{fig_flux5}. In Figs. \ref{fig_p1}, and \ref{fig_p2}, we spot-check some red stars from the earliest NGSL run, proposal 9088. Figure \ref{fig_p3} spot-checks a few stars from proposal 13776, NGSL stars that have not yet been incorporated into a higher-level science product (HLSP). Note that our scheme underestimates the scattered light for carbon star HD 158377, no doubt due to the relatively bizarre spectral shape of late-type carbon stars. About 73\% of NGSL targets currently reside at MAST in HLSP form as of the date of this publication, though the complete sample should appear there shortly (Pal et al. 2022, in preparation).

The various spectra illustrate the very large multiplier at the short-wavelength end in going from count rate to $F_\lambda$. This arises from the nearly-dead CCD response at the short wavelength end of the spectrum. Small fractions of counts per second translate into apparently-significant flux after the conversion in some cases. For red stars with no expected UV flux, it is perfectly acceptable to tune $K_0$ to zero out the short-wavelength flux on a case-by-case basis, for cosmetic purposes if nothing else. We don't do that here, of course, in order to show typical efficacy of our blanket correction scheme, but we give an example in Fig \ref{fig_p3}.

There is some volatility in the scattered light, as illustrated in a few of the panels in Figs. \ref{fig_p1} -- \ref{fig_p3}. We also find that, in the cases of HD~102212 and HD~13520, the older NGSL exposures had $~$10\% more scattered light than our 2021 observations. The cause is probably unreproducible variations in the positioning of the mode selection mechanism (MSM), which selects and retilts gratings. When the grating is shifted relative to the incoming beam, different grooves and different microscopic defects will be illuminated.

%% Subsection
\vspace{-0.3cm}
\ssubsection{Pointing-offset corrections}\label{sec:offsetcors}

\begin{figure}
   \centering
   \includegraphics[width=\columnwidth]{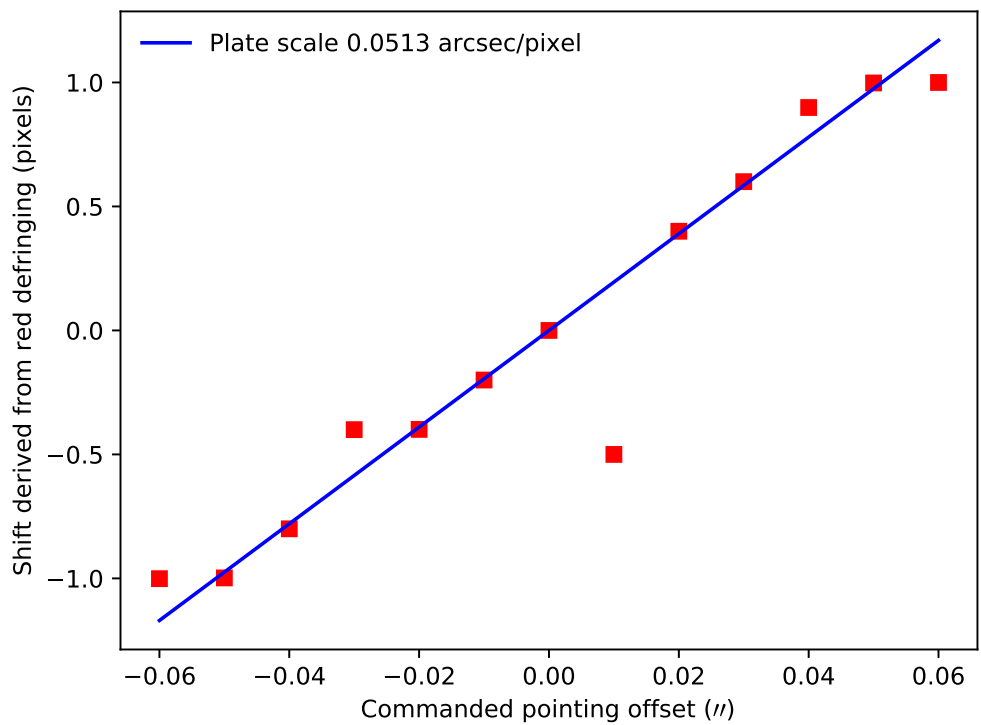}
      \caption{Offsets from slit-center derived from multiple observations of BD~+75~325, plotted against the commanded pointing offsets. 
              }
         \label{shifts}
   \end{figure}

Another systematic effect that can be partially corrected is wavelength-dependent flux attenuation (and occasional boosting) when the star does not land at the center of the 0$\farcs$2 slit. This effect was investigated with multiple observations of hot subdwarf BD~+75~325 \cite{2010AAS...21546317L}, the data for which we retrieved from the MAST archive (\url{https://dx.doi.org/10.17909/t9-3063-ma41}).  In steps of 0$\farcs$01, the star was positioned relative to slit center (at zero) in the range $-0\farcs06$ --- $0\farcs06$. If red spectra (G750L) are present in the observation set, as they are with the NGSL, one can confirm how well the star is positioned relative to the slit center by cross-correlating interference fringes in the spectrum with similar fringes in the ``fringe flat'' calibration image. The utility of such a procedure is shown in Fig. \ref{shifts}, which was done with default settings in the \textit{stistools defringe} task (python implementation).

% DOI: 10.17909/t9-3063-ma41   -- spectra from GO 11652

%----------------------------------------------------------------- 
%-----------------------------------------------------------------

\begin{figure}
   \centering
   \includegraphics[width=\columnwidth]{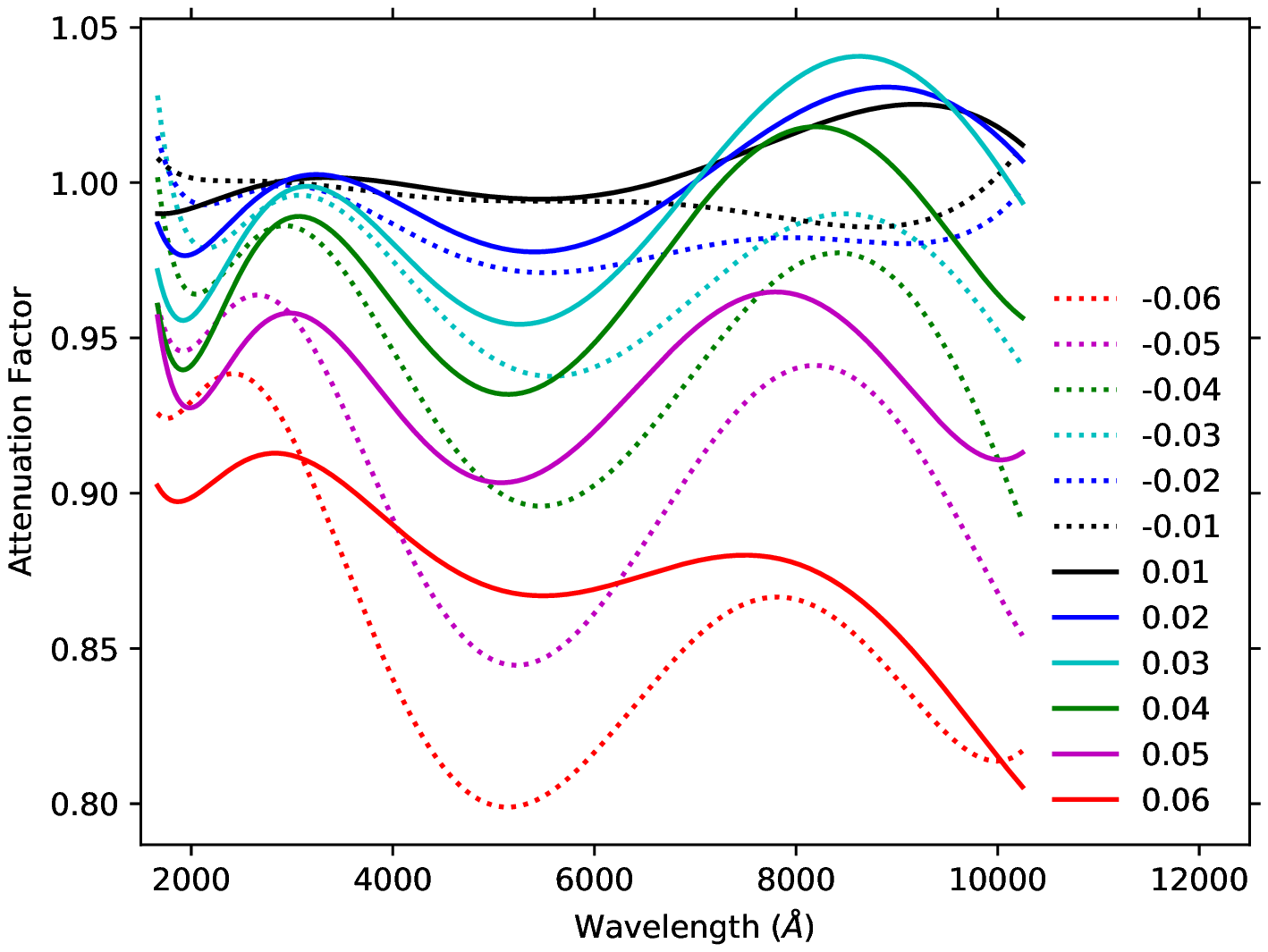}
      \caption{Corrections for point sources offset from slit center derived from multiple observations of BD~+75~325, presented as smoothed functions of wavelength. Offsets in arcseconds are as keyed. Correct by dividing the function into  the spectrum.
              }
         \label{offcenter1}
   \end{figure}

Relative to slit center, offset observations generally become fainter as more and more of the PSF is cut off by the impending slit edge. For the NGSL, the BD~+75~325 observations, and our own observations, the slit width is set at $0\farcs2$. Smoothed attenuation curves ($D_\lambda$) are shown in Figure \ref{offcenter1}. The curves themselves can be generated with the coefficients listed in Table \ref{coeffs}. Those coefficients substitute into the formula

$$  D_\lambda = a + bq + cq^2 + dq^3 + eq^4 + fq^5 + gq^6 $$

where $ q= \sqrt{\lambda / 4500\  \angstrom} $ , chosen to decrease the number of terms needed to adequately follow the twists and turns in the corrections.

%----------------------------------------------------------------- 
%-----------------------------------------------------------------
%-----------------------------------------------------------------
   \begin{table*}
      \caption[]{Slit-offcenter correction polynomial coefficients}
         \label{coeffs}
     $$ %\footnotesize
         \begin{array}{rrrrrrrr}
            \hline
            \noalign{\smallskip}
            {\rm Offset\ } (\arcsec)    &  a &  b  & c & d & e & f & g\\
            \noalign{\smallskip}
            \hline
            \noalign{\smallskip}

-0.06 & 51.7919 & -334.3680 &  891.0195 & -1230.6391 &  929.1229 & -364.0754 & 57.9600 \\
-0.05 & 63.8366 & -398.6971 & 1025.9940 & -1370.3281 & 1001.6898 & -380.3103 & 58.6770 \\
-0.04 & 62.2378 & -375.2510 &  933.2175 & -1205.1539 &  852.2159 & -313.0754 & 46.7286 \\
-0.03 & 51.9797 & -307.4814 &  754.1308 &  -962.4167 &  674.0547 & -245.8139 & 36.5035 \\
-0.02 & 29.1020 & -173.6559 &  437.9766 &  -577.0897 &  419.1689 & -159.3072 & 24.7838 \\
-0.01 &  9.5418 &  -54.2876 &  141.6155 &  -193.9473 &  147.0437 &  -58.5317 &  9.5605 \\
0.01  &  3.6207 &  -14.7231 &   31.6600 &   -32.3894 &   15.3595 &   -2.2709 & -0.2594 \\
0.02  & 25.2794 & -147.5373 &  362.4186 &  -460.8562 &  320.0924 & -115.2217 & 16.8106 \\
0.03  & 48.2160 & -291.2297 &  727.5100 &  -942.7146 &  668.7134 & -246.4419 & 36.9119 \\
0.04  & 68.5755 & -422.2492 & 1070.8983 & -1412.0620 & 1021.6252 & -384.9364 & 59.0906 \\
0.05  & 70.4925 & -436.0307 & 1113.0949 & -1480.8936 & 1083.3793 & -413.5598 & 64.4284 \\
0.06  & 27.8036 & -170.0663 &  436.1630 &  -580.6800 &  423.4523 & -160.5529 & 24.7577 \\

% updated
            \noalign{\smallskip}
            \hline
         \end{array}
     $$ 
   \end{table*}
%-----------------------------------------------------------------

Our observations also contained off-slit-center pointings to confirm the attenuation behavior when the target is not centered, but only at offsets of $\pm 0\farcs05$. If there was interplay between scattered light and pointing offset, we wished to know of it (c.f., $\S$3.4). We also included observations at $\pm 0\farcs10$ in order to explore any unexpected behavior if the star was perched right at the edge of the slit. 

HST pointings are done in two phases with STIS \cite{stis_handbook}. First is an acquisition sequence, nominally accurate to $\pm 0\farcs01$. For slits of width  $\le 0\farcs1$, a peakup is recommended to center the object within $\pm 5$\% of the overall slit width. For NGSL ($0\farcs2$ slit) this peakup was not performed. It was performed for GO 16188, but we appeared to be unlucky. Only the CCD observation of HD 13520 landed near the slit centerline. The remainder of both CCD and MAMA observations appear to have placed the star $-0\farcs02$ to $-0\farcs03$ on the negative side of center. We can tell because we requested observations placed off slit center by $\pm 0\farcs05$ and $\pm 0\farcs10$, and these observations showed asymmetrical fluxes (nearly ``full on'' for a $+0\farcs10$ displacement but closer to "full off" for the $-0\farcs10$ displacement, for example. Because we took only UV observations, it is not possible to measure this accurately by comparing the G750L red fringe pattern with fringe flat observations.

Due of these considerations, our data are less suited to determine the illumination pattern corrections than the Lindler \& Heap \cite*{2010AAS...21546317L} observations, and so we do not attempt it. Graphical comparisons indicate that our data are consistent with the BD~+75~325 observations summarized in Figure \ref{offcenter1} and Table \ref{coeffs}.

%% Subsection
\vspace{-0.3cm}
\ssubsection{Temperature of irrelevance}\label{sec:irrelevance}

Hot stars emit copiously in the UV, and this should overwhelm the scattered light contribution at some stellar temperature. At what stellar temperature does the scattered light correction become superfluous? The answer is partly wavelength-dependent. For example, if one is studying only Mg II $\lambda$2800, then scattered light is a small effect even for M stars. However, if one wishes accurate fluxing to the short wavelength edge of the wavelength coverage, then because of the large correction factors for sensitivity, small amounts of scattered light matter. 

The clearest illustration of this comes from Fig. 3 of Dashevsky \& Caldwell \cite*{2000AAS...196.3211D}, who plot G2 V star GSPC~P330-E with and without scattered light to show detectable contamination between 1650\AA\ and 2000\AA\ after flux corrections are applied. If fluxing to the short wavelength edge is a goal, then, we recommend applying the correction at least to mid-F spectral type. Stretching to the short wavelength edge is not a priority for the NGSL library, so one might relax that guideline to about G0 spectral type. For the NGSL, that is the majority of the stars. 

There is, of course, no harm done by always applying the correction, even when it is negligible in relative magnitude.

%% Subsection
\vspace{-0.3cm}
\ssubsection{Coupling between off-center observations and scattering}\label{sec:coupling}

The effects of off-slit-center observations on scattered light can be gauged by scaling fully-illuminated observations (on the slit centerline) down to match substantially attenuated versions of the same observation (close to the edge of the slit). If scattering is independent of target placement relative to the centerline then the relative count rates should match, within the noise. The more direct scheme of isolating the scattered light by subtracting MAMA-based spectra is somewhat hampered since the MAMA observations have pointing offsets relative to the CCD observations. In any case, all comparisons indicate that the ratio of stellar and scattered light is a constant. Attenuation of the stellar signal produces a proportional attenuation of the scattered light signal. 

The only remaining calibration issue is therefore reducing the pedestal level $K_0$ whenever observations are off-slit-center enough to cause significant attenuation. If $\pm$5\% is taken as a suitable threshold for concern, then the star needs to be within $\pm 0\farcs04$ of center, according to Fig. \ref{offcenter1}. Or, one can simply apply the attenuations from Table \ref{coeffs} at approximately $\lambda$5450, apply that correction to the $V$ pseudoflux, and calculate $K_0$ from the formula in Table \ref{ktab}.

All of these issues regarding centering refer to the $0\farcs2$ slit. In a wider slit the object acquisition is accurate enough to place the point source far enough away from slit edges to mitigate edge effects. And for a smooth extended source the slit should be uniformly illuminated across its width and therefore the illumination pattern arriving at the detector should be the straightforward convolution of the PSF with the slit.

Given the constancy of the scattered light as a function of placement relative to slit center, a correction can be derived for the case of an extended source that uniformly illuminates the slit. The convolution is over slit position and also the spatial dimension, meaning that the railroad track spectra must also be included. This was done, yielding, for extended sources 

$$ S(\lambda) = L_0 [1.0 + 8.27 \times 10^{-4}*(\lambda - 2000)],$$

where $L_0$ is the amount of light at $\lambda$2000 for an extended source. Because the railroad track spectra are less dependent on wavelength, the slope of this relation is flatter. Relations of count rate $L_0$ to fluxes are given in Table \ref{ltab}. A magnitude used for finding $L_0$ should be the expected magnitude for the spatial area of the object over the extraction region and slit width. This formula and the Table \ref{ltab} scalings are for stellar or galactic continuous spectra. However, if the extended spectrum is dominated by emission lines, one might arrive at a reasonable guess for $L_0$ by estimating the count rate $C$ expected from O III $\lambda$5007 plus any other strong lines in the vicinity of the $V$ bandpass.

\begin{table}
\caption{Estimators of $L_0$, the count rate at $\lambda$2000 for extended sources, with fractional absolute deviations for slope $A$.}             % title of Table
\label{ltab}      % is used to refer this table in the text
\centering                          % used for centering table
\begin{tabular}{lll}        % centered columns (4 columns)
\hline\hline                 % inserts double horizontal lines
 title & formula & $\Delta A / A$ \\    % table heading 
\hline                        % inserts single horizontal line
   $L_0 =$ & $594 \times 10^{-0.4\ V} $ & 0.01 \\   % body of the table
   $L_0 =$ & $2366 \times 10^{-0.4\ B} $ &  0.02\\ 
   $L_0 =$ & $6.71 \times 10^{-7}  \times\ C$ & 0.02  \\
   $L_0 =$ & $417 \times 5 \times 10^8 \times F_{5000}$ & 0.08 \\
   $L_0 =$ & $269 \times 5 \times 10^8 \times F_{6000}$ & 0.10 \\
   $L_0 =$ & $2.02 \times 10^7 \times\ F_\lambda$ & 0.10 \\
   $L_0 =$ & $280 \times 10^{-0.4\ G} $ & 0.18 \\ 
\hline    

\end{tabular}
\end{table}

%% Subsection
\vspace{-0.3cm}
\ssubsection{Standalone G230LB data}\label{sec:standalone}

In the absence of data beyond the G230LB spectrum for stellar sources a scheme to roughly estimate the amount of red scattered light is still possible. To attempt a model for this, we used stellar spectra obtained in proposal numbers GO 9088, GO 9786, GO 10222, and GO 13776 that comprise raw NGSL observations. These $\sim$540 stellar targets cover all temperatures. The observational data were supplemented with $V$ magnitudes from SIMBAD. The idea is to extrapolate from the UV portion of the spectrum to predict $V,$ and then use Table \ref{ktab} to estimate the strength of the correction.

\begin{figure}
   \centering
   \includegraphics[width=\columnwidth]{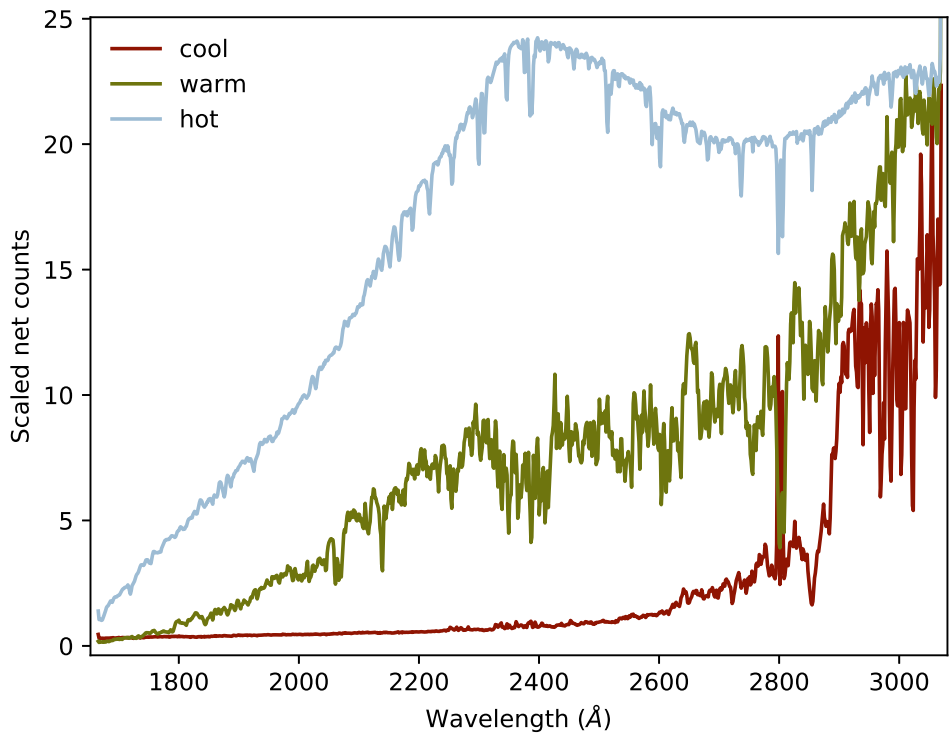}
      \caption{A hot, warm, and cool star as seen by STIS/G230LB before fluxing. These are HD 86986 (A1 V), HD 89388 (K2.5 II), and HD 30834 (K3 III), respectively. The predictor formula attempts to predict $V$ flux at $\lambda$5450 based on UV flux at $\lambda$2900 and convex/concave shape.
              }
         \label{b2examp}
   \end{figure}

The $V$ flux extrapolation starts with the UV flux at $\lambda$2900 and adds convex/concave shape information as detailed here. Example stars as seen through the instrument are shown in Fig. \ref{b2examp}. After extraction, each spectrum was passed through a median filter to reject noise spikes, namely scipy.ndimage's median\_filter with size = 5. Effects of metallicity, surface gravity, or dust on the continuum shape were neglected. We collect three bins of average count rate: $f_{20}$ = 2000 $\pm$100\AA, $f_{24}$ = 2450 $\pm$100 \AA, and $f_{29}$ = 2975 $\pm$75 \AA. Very late-type stars with near-zero mid-UV flux can be effectively isolated by requiring $f_{24}/f_{20} < 2.15$ (these bins will contain only scattered light). A threshold much higher than 2.15 can include very hot stars, which also have a relatively flat mid-UV spectrum. For very late-type stars, one should simply adopt the measured $f_{20}$ as $K_0$ in the correction formula.

\begin{figure}
   \centering
   \includegraphics[width=\columnwidth]{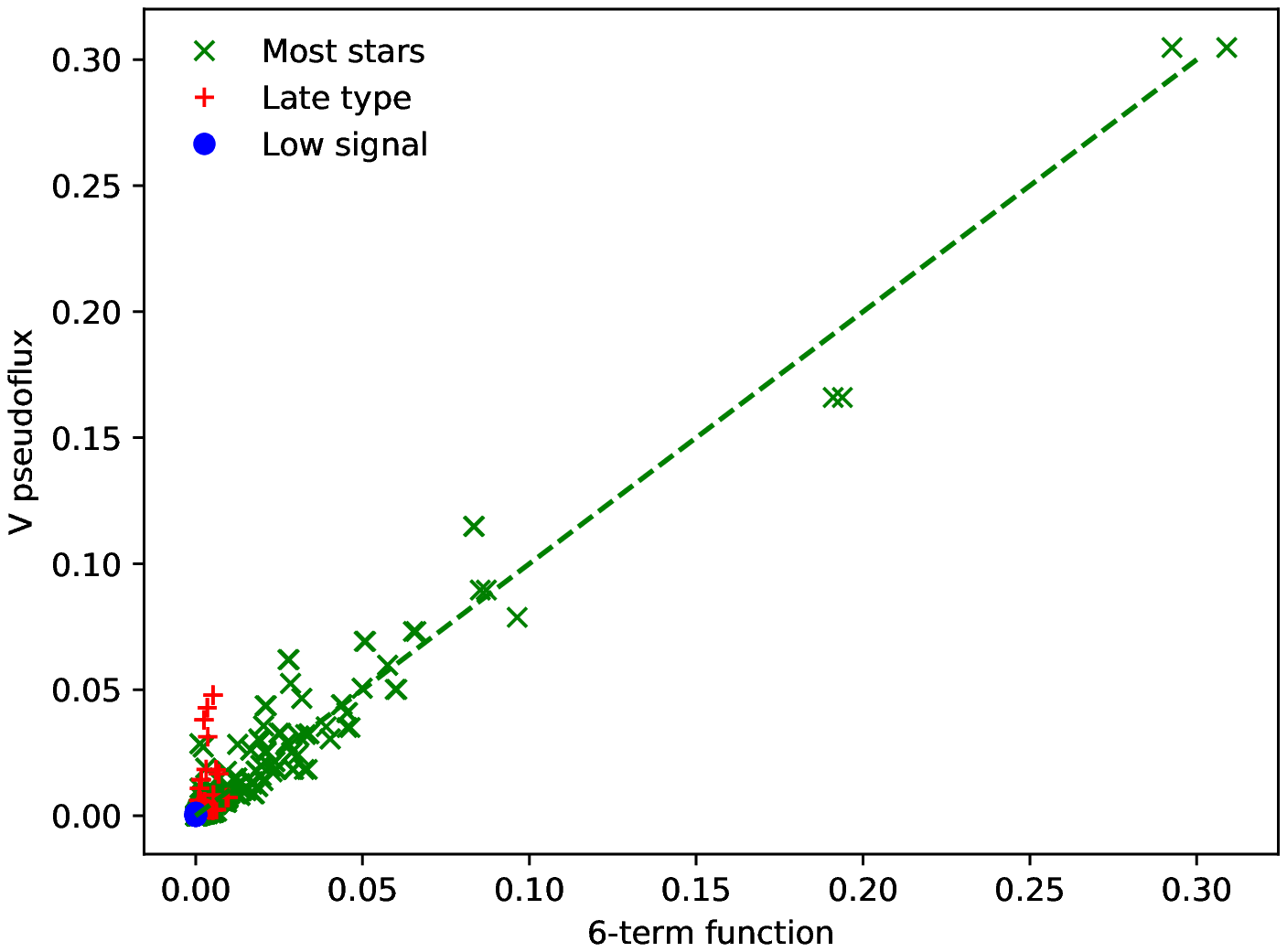}
      \caption{A 6-coefficient fit to UV passbands to attempt to predict $V$ pseudoflux compared to measured $V$ pseudoflux. Dashed line is $x=y$. Late-type stars (red crosses) and stars with low signal (blue dots) are excluded from the fit, leaving warm and hot stars (green tilted crosses). The result is approximate, and gets worse, in a fractional sense, for pseudofluxes less than 0.01.
              }
         \label{bohlin2}
   \end{figure}

The remaining warm and hot stars were fit with

$$ P_V = a*f_{29} ( b*Y + c*Y^2 + d*Z + e*Z^2 + f*Z^3 ) $$

where $Y = f_{29}/f_{24}$ is a red slope term and $Z = \frac{1}{2}(f_{20} + f_{29})/f_{24}$ is a curvature term.  $P_V$ refers, not to the standard magnitude, but to the pseudoflux $P_V = 10^{-0.4 V}$. The correction function becomes  $ S(\lambda) = 426 P_V [1.0 + 0.00104*(\lambda - 2000)]$. The coefficients are listed in Table \ref{functab} and an idea of how well the formula predicts $V$ pseudoflux can be seen in Fig. \ref{bohlin2}.

\begin{table}
\caption{Fit coefficients for UV spectra prediction of $V$ pseudoflux.}    % title of Table
\label{functab}      % is used to refer this table in the text
\centering                          % used for centering table
\begin{tabular}{ll}        % centered columns (4 columns)
\hline\hline                 % inserts double horizontal lines
 coefficient & value \\    % table heading 
\hline                        % inserts single horizontal line
   a & $9.6186 \times 10^{-7}$ \\      % inserting body of the table
   b & 3.9942 \\
   c & $-$2.3103 \\
   d & $-$7.3421 \\
   e & 9.3143 \\
   f & $-9.7158 \times 10^{-3}$ \\ 
\hline    

\end{tabular}
\end{table}

%----------------------------------------------------------------- 
%-----------------------------------------------------------------

%----------------------------------------------------------------- 
%-----------------------------------------------------------------

\vspace{-0.3cm}
\ssubsection{Cookbook}\label{sec:cookbook}

\begin{enumerate}
\item{For a stellar source, find the $V$ magnitude, and compute $K_0 = 426 \times 10^{-0.4 V}$.}
\item{From the standard extraction package x1d output, read both NET counts and FLUX. If ATODGAIN is not equal to one, multiply the NET counts by the gain. Divide FLUX by NET for a convenient correction spectrum, already correctly binned.}
\item{From the NET counts spectrum, subtract $S(\lambda) = K_0 [1.0 + 0.00104*(\lambda - 2000\ \angstrom)] $. }
\item{Correct back to flux with the correction spectrum from step 2.}
\end{enumerate}

%% SECTION - CONCLUSIONS
\vspace{-0.3cm}
\ssection{Conclusions}\label{sec:conclusions}

We investigated scattered light in the G230LB grating of the STIS instrument aboard HST. 

   \begin{enumerate}
      \item Red stars show substantial scattered red light that resembles signal throughout the UV.
      \item The scattered light is modeled as a gently sloping ramp. The wavelength-dependent ramp for count rate is given by $$ S(\lambda) = K_0 [1.0 + 0.00104*(\lambda - 2000\ \angstrom)],$$ where $K_0$ is the count rate of scattered light at $\lambda2000$.
      \item If the $V$ magnitude is known, $K_0 = 426 \times 10^{-0.4\ V}$ counts per second, and formulae are given for other flux estimators as well. Fluxes from passbands that too far redward or blueward from $V$ do not predict $K_0$ very well. An approximate formula is presented for the case where only G230LB data are available.
      \item For the special case of the $0\farcs2$ slit, wavelength-dependent flux corrections are developed for stars which land off the center of the long slit. These corrections are based on one set of calibration images and are approximate, as they do not include focus-dependent PSF changes that other observation sets might encounter. The corrections are given in the form of tabulated polynomial coefficients.
      \item For stars warmer than about G0 spectral type, the scattered light contribution becomes optional.
      \item Scattered light appears not to couple with the placement of the star relative to slit center. This allows a straightforward extrapolation to the case of an extended source that spans the slit. $$ S(\lambda) = L_0 [1.0 + 0.000827*(\lambda - 2000\ \angstrom)],$$ where $L_0$ is the count rate of scattered light at $\lambda2000$. $L_0$ is somewhat larger than $K_0$.
      \item The ``railroad tracks'' (STIS handbook $\S$13.7.4; ISR-98-24) are composed of scattered light. They only appear for very red stars.
   \end{enumerate}

The scattering primarily arises in the grating itself, but the presence of railroad track spectra imply the presence of additional internal reflection. By themselves, the data we present cannot delve more specifically into which optical components might be involved (besides the G230LB grating). The amount of scattered light varies by several tens of percent compared to the formula, plausibly due to MSM positioning shifts.

The formulae presented here should find use in a fair fraction of the more than 3000 observations taken with the G230LB grating. They provide an observation-based scattered light correction for yellow and red stars observed at low resolution in the UV, with accompanying increases in flux and line strength accuracy.

%%% ACKNOWLEDGEMENTS %%% 
\vspace{-0.3cm}
\ssectionstar{Acknowledgements}
\vspace{-0.3cm}

Based on observations made with the NASA/ESA Hubble Space Telescope, program GO 16188, \url{https://dx.doi.org/10.17909/t9-d42d-z465}. Support for this work was provided by NASA through grant number HST-GO-16188.001-A from the Space Telescope Science Institute. STScI is operated by the Association of Universities for Research in Astronomy, Inc. under NASA contract NAS 5-26555. We would like to thank Dr. D. Welty for his excellent comments.

%%% CHANGE HISTORY %%%
\vspace{-0.3cm}
%Put instrument, year, and ISR number
\ssectionstar{Change History for STIS ISR 2022-05}\label{sec:History}
\vspace{-0.3cm}
Version 1: \ddmonthyyyy\today - Original Document %Month DD, YYYY format

\bibliographystyle{astron}
\bibliography{g230lb} 

\begin{thebibliography}{}

\bibitem[\protect\astroncite{{Apai} et~al.}{2010}]{2010PASP..122..808A}
{Apai}, D., {Lagerstrom}, J., {Reid}, I.~N., {Levay}, K.~L., {Fraser}, E.,
  {Nota}, A., and {Henneken}, E.: 2010,
\newblock {\em PASP} {\bf 122(893)}, 808

\bibitem[\protect\astroncite{Branton et~al.}{2021}]{stis_handbook}
Branton, D., Riley, A., et~al.: 2021,
\newblock {\em STIS Instrument Handbook, Version 20.0},
\newblock STScI, Baltimore

\bibitem[\protect\astroncite{{Dashevsky} and
  {Caldwell}}{2000}]{2000AAS...196.3211D}
{Dashevsky}, I. and {Caldwell}, J.: 2000,
\newblock in {\em American Astronomical Society Meeting Abstracts \#196}, Vol.
  196 of {\em American Astronomical Society Meeting Abstracts}, p. 32.11

\bibitem[\protect\astroncite{{Gregg} et~al.}{2004}]{2004AAS...205.9406G}
{Gregg}, M.~D., {Silva}, D., {Rayner}, J., {Valdes}, F., {Worthey}, G.,
  {Pickles}, A., {Rose}, J.~A., {Vacca}, W., and {Carney}, B.: 2004,
\newblock in {\em American Astronomical Society Meeting Abstracts}, Vol. 205 of
  {\em American Astronomical Society Meeting Abstracts}, p. 94.06

\bibitem[\protect\astroncite{{Heap} and {Lindler}}{2008}]{2008AAS...21116225H}
{Heap}, S.~R. and {Lindler}, D.: 2008,
\newblock in {\em American Astronomical Society Meeting Abstracts \#211}, Vol.
  211 of {\em American Astronomical Society Meeting Abstracts}, p. 162.25

\bibitem[\protect\astroncite{Heap and Lindler}{2010}]{MASTNGSL}
Heap, S.~R. and Lindler, D.: 2010,
\newblock {\em STIS Next Generation Spectral Library, Version 2, based on AR
  11755, GO 11652, March 2010},
\newblock \url{https://archive.stsci.edu/prepds/stisngsl/}

\bibitem[\protect\astroncite{{Heap} and {Lindler}}{2016}]{2016ASPC..503..211H}
{Heap}, S.~R. and {Lindler}, D.: 2016,
\newblock in S. {Deustua}, S. {Allam}, D. {Tucker}, and J.~A. {Smith} (eds.),
  {\em The Science of Calibration}, Vol. 503 of {\em Astronomical Society of
  the Pacific Conference Series}, p. 211

\bibitem[\protect\astroncite{{Lindler} and {Heap}}{2010}]{2010AAS...21546317L}
{Lindler}, D. and {Heap}, S.: 2010,
\newblock in {\em American Astronomical Society Meeting Abstracts \#215}, Vol.
  215 of {\em American Astronomical Society Meeting Abstracts}, p. 463.17

\bibitem[\protect\astroncite{Sohn et~al.}{2019}]{data_handbook}
Sohn, S.~T. et~al.: 2019,
\newblock {\em STIS Data Handbook, Version 7.0},
\newblock STScI, Baltimore

\bibitem[\protect\astroncite{{Woodgate} et~al.}{1998}]{1998PASP..110.1183W}
{Woodgate}, B.~E., {Kimble}, R.~A., {Bowers}, C.~W., {Kraemer}, S., {Kaiser},
  M.~E., {Danks}, A.~C., {Grady}, J.~F., {Loiacono}, J.~J., {Brumfield}, M.,
  {Feinberg}, L., {Gull}, T.~R., {Heap}, S.~R., {Maran}, S.~P., {Lindler}, D.,
  {Hood}, D., {Meyer}, W., {Vanhouten}, C., {Argabright}, V., {Franka}, S.,
  {Bybee}, R., {Dorn}, D., {Bottema}, M., {Woodruff}, R., {Michika}, D.,
  {Sullivan}, J., {Hetlinger}, J., {Ludtke}, C., {Stocker}, R., {Delamere}, A.,
  {Rose}, D., {Becker}, I., {Garner}, H., {Timothy}, J.~G., {Blouke}, M.,
  {Joseph}, C.~L., {Hartig}, G., {Green}, R.~F., {Jenkins}, E.~B., {Linsky},
  J.~L., {Hutchings}, J.~B., {Moos}, H.~W., {Boggess}, A., {Roesler}, F., and
  {Weistrop}, D.: 1998,
\newblock {\em PASP} {\bf 110(752)}, 1183

\end{thebibliography}

\end{document}